%% file: main.tex
\documentclass[conference]{IEEEtran}
\usepackage[colorlinks,urlcolor=blue,linkcolor=blue,citecolor=blue]{hyperref}

\usepackage{graphicx} 

\usepackage[table]{xcolor} 
\usepackage{amssymb} 
\usepackage{soul} 
\usepackage{tabularray}
\usepackage{ragged2e}
\usepackage{booktabs}
\usepackage{multirow}
\usepackage{amsmath}

\usepackage{algorithm}
\usepackage{algorithmicx}
\usepackage{algpseudocode}

\newcommand{\grey}{\cellcolor[HTML]{E8E8E8}}

\title{parHSOM: A novel parallel Hierarchical Self-Organizing Map implementation}

\author{
\IEEEauthorblockN{Rebekah Lane}
\IEEEauthorblockA{\textit{Computer Science \& Engineering} \\
\textit{Mississippi State University}\\
Starkville, United States \\
rel250@msstate.edu}
\and
\IEEEauthorblockN{Logan Cummins}
\IEEEauthorblockA{\textit{Computer Science \& Engineering} \\
\textit{Mississippi State University} \\
Starkville, United States \\
nlc123@msstate.edu}
\and
\IEEEauthorblockN{Andy Perkins}
\IEEEauthorblockA{\textit{Computer Science \& Engineering} \\
\textit{Mississippi State University} \\
Starkville, United States \\
perkins@cse.msstate.edu}
\and
\IEEEauthorblockN{George Trawick}
\IEEEauthorblockA{\textit{Computer Science \& Engineering} \\
\textit{Mississippi State University} \\
Starkville, United States \\
gtrawick@cse.msstate.edu}
\and
\IEEEauthorblockN{Ioana Banicescu}
\IEEEauthorblockA{\textit{Computer Science \& Engineering} \\
\textit{Mississippi State University} \\
Starkville, United States \\
ioana@cse.msstate.edu}
\and
\IEEEauthorblockN{Sudip Mittal}
\IEEEauthorblockA{\textit{Computer Science} \\
\textit{University of Alabama} \\
Tuscaloosa, United States \\
sudip.mittal@ua.edu}
}

\begin{document}

\maketitle

\begin{abstract}
    The digital age has completely transformed the way that information is processed and stored, which makes cybersecurity a crucial field of research. Cybersecurity contains many different domains, but this work focuses on Intrusion Detection Systems (IDSs). Within the literature, Hierarchical Self-Organizing Maps (HSOMs) have been used to create trustworthy, explainable, and AI-based IDSs. However, HSOMs are trained sequentially, which means that training HSOMs on large datasets is slow. This work presents a novel parallel HSOM architecture, called parHSOM. The purpose of this research is to investigate the effect that parallel computation has on the HSOM training time. parHSOM is tested on two different testbeds, four different output grid sizes, and five different cybersecurity datasets. Performance metrics collected from these experiments show that parHSOM consistently trains faster than the Sequential HSOM algorithm without any significant loss in performance. Additionally, this work provides a platform for further investigation into parallel HSOM implementations.
\end{abstract}


\section{Introduction} \label{intro}
The rise of the digital age has completely transformed the way information is processed and stored. Digital data provides greater availability compared to physical, but the cost of availability is an increased risk of exploitation. Protecting devices, systems, and their accompanying digital data is the priority of cybersecurity. Cybersecurity contains many different domains, but this work focuses on Intrusion Detection Systems (IDS). Cyberattacks are continuously changing, and thus, researchers are constantly studying various ways to improve IDS. Previous literature, has shown that artificial intelligence (AI) can be used to improve the quality of IDS. However, in order for AI-based IDS to be beneficial, cybersecurity analysts need to be able to trust their model's decision making process. Explainable AI (XAI) is a field of research that explores methods for improving AI trustworthiness by providing explanations about a model's decision making process. Operators are then able to use these explanations to understand how the AI functions.

Self-Organizing Maps (SOMs), a type of XAI that clusters data, are commonly used to create understandable, AI-based IDSs \cite{ables2022creating}. The SOM algorithm creates a 2D graph representations of the relationships between data points, which provides explanations for why data is being clustered a particular way. However, the SOM algorithm is trained sequentially and is sensitive to initial conditions \cite{cuello2024SOM}.

The Hierarchical Self-Organizing Map (HSOM) is an extension of the SOM that was designed to mitigate the SOM's sensitivity to initial conditions. The HSOM provides greater clustering detail than the SOM, and it inherits the SOM's advantages of trustworthiness and explainability. The HSOM is also trained sequentially, which quickly becomes impractical for large datasets. One solution to the slow training time of the HSOM would be to utilize parallel computation.

To the best of our knowledge, a parallel HSOM has not been implemented. Therefore, the purpose of this work is to design parHSOM, a parallel implementation of the HSOM algorithm. Parallel computation could enable the HSOM to train faster; moreover, the HSOM can be trained on more data resulting in a more robust model. This work analyzes various methods used to parallelize the SOM algorithm and uses that knowledge as a foundation for parHSOM. The purpose is to answer two important research questions:
\begin{itemize}
    \item [RQ1:] How can the Hierarchical Self-Organizing Map (HSOM) algorithm be parallelized?
    \item [RQ2:] What is the performance of the parallel Hierarchical Self-Organizing Map (parHSOM) algorithm in comparison to the sequential HSOM algorithm?
\end{itemize}
The model proposed in this research parallelizes the HSOM training process and has been tested extensively on two different test environments, multiple different output grid sizes, and five different cybersecurity datasets that are prominent in the literature. To provide a thorough analysis of the parHSOM performance and to provide an answer to RQ2, this work collected the following performance metrics - accuracy, precision, F1 Score, False Positive Rate (FPR), False Negative Rate (FNR), Training Time (TT), and Prediction Time (PT). Through experimentation, this work found that parHSOM consistently trains faster than the Sequential HSOM, with a maximum speed increase of 6.056 times faster than the Sequential HSOM. Furthermore, parHSOM performs similarly to the Sequential HSOM in terms of accuracy, precision, F1 score, FPR, FNR, and PT.

The layout for this work is as follows - Section \ref{background} provides more detail on the SOM and HSOM and defines how this work researched different methods for parallelizing the HSOM, Section \ref{design} outlines the design for parHSOM and discusses how the experiments for this research were conducted, Section \ref{results} provides the results from the experiments, Section \ref{discussion} discusses how the experiment results relate to RQ1 and RQ2, and Section \ref{conclusion} summarizes the findings of this research, lists limitations of this work, and provides future areas of research.

\section{Background \& Related Works} \label{background}
\subsection{Self-Organizing Maps (SOMs)}
Self-Organizing Maps (SOMs) have been used in numerous fields of research, such as wireless networks, Internet of Things (IoT), and image processing \cite{panwar2024data, bagher2021edgesom, de2016parallel}. The key advantage of using a SOM is that they have the ability to provide dimensionality reduction without loss of topographical information. Additionally, SOM networks are unsupervised and trained using competitive learning. Combined together, these features create an explainable AI model that is especially useful in the field of cybersecurity. The authors of \cite{ables2022creating} support this reasoning by demonstrating how SOMs can be used to create reliable, trustworthy Intrusion Detection Systems (IDS).

The architecture of the SOM network is simple, consisting of only two layers --- an input layer and an output grid. It is trained using competitive learning, and thus, does not require labeled data. As data samples are fed into the SOM, the output neurons are modified to detect patterns and clusters in the training dataset \cite{kohonen1990self}. The output of a SOM network can be utilized to generate two dimensional (2D) graph representations of relationships between the input samples.

\subsection{SOM Algorithm}
This section discusses the various equations involved in the SOM training process, which have been provided by the authors of \cite{mancini2020xpysom}. The input layer of the SOM algorithm takes in data samples from the training dataset. The dataset can be represented by the following equation:
\begin{equation} \label{eq:nbr1}
    \mathbb{X}=\{x_0,\;x_1,...,\;x_{N-1}\}\;\; x_i \in \mathbb{R}^P,\;\forall i=0...N-1
\end{equation}


\noindent where $x_i$ represents the individual dataset samples. The output grid is connected to the input layer utilizing weights, which are randomly initialized at the beginning of the training process. These connections are represented by equation \ref{eq:nbr2}.

\begin{eqnarray} \label{eq:nbr2}
    \mathbb{W}=\{w_0,w_1,...,w_{M-1}\}\;\; w_k \in \mathbb{R}^P,\; \forall k=0..M-1
    \nonumber\\
    where\;M=G_WG_H,\; arranged \; as \; a \; G_W \times G_H \; grid
\end{eqnarray}

As mentioned previously, SOMs are used to cluster data. The first step is to randomly select a data sample from the training dataset. Then, the Best Matching Unit (BMU) is calculated for that sample using the following equation:
\begin{equation} \label{eq:nbr3}
    b_i=argmin_k|x_i-w_k|_2
\end{equation}
\noindent The BMU specifies which neuron in the output grid is most similar to the input sample. The goal is to adjust the BMU, and its surrounding neighbors, so that it becomes more similar to the input sample. Adjusting the neurons in the output layer slowly fits the grid to our training data, which allows the clusters to form. The neighborhood function is calculated with equation \ref{eq:nbr4}:
\begin{equation} \label{eq:nbr4}
    h(b,k,t)=-exp \left(\frac{||r_b-r_k||^2}{\delta(t)}\right)
\end{equation}

The presence of the neighborhood function is what differentiates the SOM algorithm from the K-means clustering algorithm \cite{cui2023selforganizing}. Instead of updating only one output neuron per input sample, the neighborhood function allows us to also update the neurons surrounding the BMU, albeit to a lesser degree.

Once the neighborhood function has been defined, the necessary updates for each specified weight can be calculated using equation \ref{eq:nbr5}:
\begin{equation} \label{eq:nbr5}
    w_k(t{'}\text{+}1)=w_k(t{'})+\alpha(t)h(b_i,k,t)(x_i-w_k(t{'}))
\end{equation}
\noindent During the SOM training process, data samples are randomly selected, the BMU and neighborhood function are calculated, and the weights are updated until the convergence criteria, which are set by the researcher, have been met.

In the following subsection, this work will discuss the advantages and the disadvantages of the SOM.

\subsection{Advantages and Disadvantages of the SOM Algorithm}
The three main advantages of the Self-Organizing Map (SOM) that this paper focuses on are dimensionality reduction, topological preservation, and simplicity \cite{cuello2024SOM, ghorpade2023tutorial}. Dimensionality reduction refers to the SOM property where information from a high dimensional dataset is mapped onto a 2D output grid. The output grid can then be used to generate a 2D graph. Topological preservation refers to the fact that no important information is lost during dimensionality reduction \cite{ghorpade2023tutorial}. Finally, the SOM algorithm is simple and easy to understand because it is based off of the Euclidean distance equation. Combined together, these advantages result in a trustworthy and explainable AI algorithm.

However, the SOM also has disadvantages \cite{cuello2024SOM}, and this paper focuses on two of them. First, it is trained sequentially. Thus, as the size of the training dataset is increased, so does the SOM training time. Second, the initialization variables for a SOM will affect the output of that SOM, which makes it sensitive to initial conditions.

\subsection{Hierarchical Self-Organizing Maps (HSOMs)}\label{hsom}
\begin{figure}[!ht]
    \centering
    \includegraphics[width=.5\linewidth]{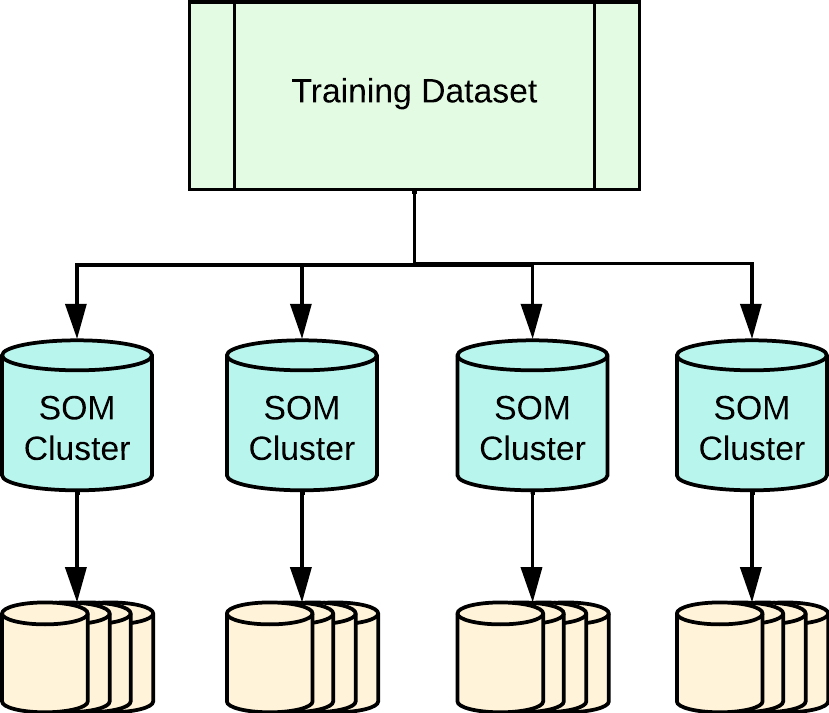}
    \label{fig_hsom_model}
    \caption{A visual representation of the output from the HSOM algorithm.}
\end{figure}

The Hierarchical Self-Organizing Map (HSOM) is an extension of the SOM algorithm. It is unsupervised, uses competitive learning, and it inherits the advantages and disadvantages of the SOM. However, unlike the SOM algorithm, the HSOM completes multiple levels of clustering, which provides greater than the SOM. The HSOM was designed to mitigate some of the SOM algorithm's sensitivity to initial conditions. It functions by training a SOM on the initial dataset. Then, it uses the resulting clusters to train a new SOM on each cluster, thereby providing more detail about the original SOM clusters. This process is repeated until training new SOMs no longer provides relevant information. The output from the HSOM algorithm is a hierarchy of clusters, where each new level supplies additional information as shown in Figure \ref{fig_hsom_model}.

As mentioned previously, the HSOM algorithm inherits dimensionality reduction, topological preservation, and simplicity from the SOM algorithm. Thus, like the SOM algorithm, the HSOM is explainable and trustworthy. However, it also inherits the SOM algorithm's slow training time, which is compounded by the fact that the HSOM algorithm trains multiple SOMs per model.

One method for reducing the training time of the HSOM model would be to decrease the size of the training dataset. However, this could reduce the robustness of the HSOM model. Another method for reducing the training time of the HSOM model would be to use parallel computation. As a result, a literature survey was conducted to see what methods are being used to parallelize the HSOM algorithm.

\subsection{Literature Survey} \label{lit_survey}
An investigation into parallel HSOM implementations within the literature returned inconclusive results, and thus, this research expanded the search to include parallelization methods for SOMs. Within the literature, SOM parallelization methods can be categorized into two main groups --- those that use software optimizations to decrease the training time and those that use hardware optimizations to decrease the training time. Hardware optimizations often require special computer architectures that are not widely available. Thus, in order to produce a more accessible model, this work focused on analyzing software optimization methods.

According to the literature, software optimizations can be divided into three categories --- data partitioned \cite{panwar2024data, phan2017dsom, kim2015distributed, vrusias2007distributing, mancini2020xpysom, bagher2021edgesom, gorgonio2008parallel, gorgonio2008combining, jayaratne2017apache, jayaratne2021unsupervised}, network partitioned \cite{hammond2006parallel, ozdzynski2002parallel, rauber2000parsom, bandeira1998training, huiwei1997parallel, myklebust1995parallel, hamalainen1997mapping, cui2023selforganizing, VojáČek2013scalable, chang2002efficient, zhu2009self, khan2013design, de2016parallel, moraes2012parallel, qu2005motion, wang2015massively}, and hybrid partitioned \cite{richardson2015extending, yang1999data}. The authors of \cite{richardson2015extending} describe network partitioned methods as techniques that break up the SOM grid, while data partitioned methods are techniques that break up the dataset. Hybrid partitioned methods are a combination of data and network partitioning methods. Once either the SOM grid or the dataset is divided into separate pieces, parallel computation can commence.

The results from the survey showed a research gap surrounding parallel HSOM algorithms thus leading to RQ1 and RQ2 described in Section \ref{intro}. 
As parallel computation provides the opportunity to create more robust HSOM-based IDSs, the goal of this work was to design and evaluate a parallel HSOM.

\section{Design \& Implementation} \label{design}
Observing the HSOM showed that it partitioned training data into independent subsets, just like the data partitioning method discussed in Section \ref{background}. Thus, this work used the data partitioning method as the foundation for parHSOM. Utilizing data partitioning should decrease the training time of each layer, and thus, decrease the overall training time of the HSOM. parHSOM only parallelizes the HSOM training process. The prediction process for the HSOM alogirthm remains unchanged. The DBGHSOM code from \cite{ables2023explainable} was used as the basis for the Sequential HSOM and a starting point for parHSOM. Since the literature survey did not reveal any parallel HSOM implementations, this work focused on starting with the simplest HSOM implementation. Thus, DBGHSOM was reduced to its most basic HSOM components to create the baseline Sequential HSOM. Then, the Sequential HSOM was parallelized using the data-partitioned method to create parHSOM.

\subsection{parHSOM Overview}
The parHSOM is designed to be trained in two phases - an initial clustering, which is sequential, and a parallel portion. In Phase 1, a SOM is trained on the initial dataset. The resulting clusters are independent of one another. In Phase 2, a child process can be spawned for each cluster to train a SOM on that cluster, which allows this section of training to be parallelized. The CPU is responsible for assigning the child processes to different CPU processors in the most efficient manner. Increasing the number of CPU processing cores, increases the amount of resources that the CPU can assign to the children processes, and thus, should decrease the overall training time of parHSOM. This system is repeated until the finishing criteria for the HSOM are met.

\subsection{The parHSOM Algorithm}
The datasets for these experiments were preprocessed in two steps. First, the dataset is normalized using the ‘Normalizer’ function from the \textit{sci-kit learn} python library. Then, it is split into training data and test data using the ‘train\_test\_split’ function from the same library. For this research, the initial dataset was split so that 80\% of the initial dataset was used for the training data and 20\% of the initial dataset was used for the test data. The parameters for the ‘train\_test\_split’ function were set so that the Sequential HSOM and parHSOM received the same training and test data for each experiment.

The first training phase of parHSOM clusters the training data provided by \textit{Preprocessing} into independent subsets and is sequential. This is accomplished by training a SOM on the initial training data. The results from this SOM are stored in SOM\_tmp\_list for further analysis in Phase 2. 

In Phase 2, the parHSOM parent process analyzes each node in SOM\_tmp\_list. If a node contains a SOM object, this means that further clustering might be necessary, and a copy of the SOM object’s information is stored in the parent\_SOM variable. Then, the parent process spawns a child process to run the ‘Vertical Growth Function’ on parent\_SOM. If a node does not contain a SOM object, then that signals the parent process that no further training is necessary. After a node is analyzed by the parent process, it is removed from SOM\_tmp\_list and stored in SOM\_list, which contains all of the information for the final HSOM model. The parent process continues to analyze nodes, spawn processes, and store the nodes until SOM\_tmp\_list is empty. Once this happens, the parent process waits for all of the child processes to finish. Then, the parent process collects the results from the child processes and adds them to SOM\_tmp\_list.

\noindent \textbf{Parallel Portion:} This model was programmed in Python using the Python Multiprocessing library. The Multiprocessing library provides backend support that handles race conditions, prevents deadlocks, and oversees synchronization of the processes and access to resources. A Multiprocessing Manager is used by the parent and child processes to share results back and forth. The Manager oversees a shared memory dictionary that allows the child processes to send back the SOMs that they trained to the parent process.

\begin{algorithm}
    \label{phase2}
    \caption{Phase 2}
    \begin{algorithmic}[1]
        \While {True}
            \While {$SOM\_tmp\_list \not=0$}
                \State $nodes \gets SOM\_tmp\_list[0]$
                \For {$node \in nodes$}
                    \If {$node == SOM\,object$}
                        \State $parent\_SOM\ \gets node\,data$
                        \State Spawn child process 
                        \State Run $vertical\_growth(parent\_SOM)$
                    \EndIf
                \EndFor
                \State Append $SOM\_tmp\_list[0] \rightarrow SOM\_list$
                \State Pop $SOM\_tmp\_list[0]$
            \EndWhile
            \State Wait on child processes to finish
            \State Retrieve $child\_process\_info$
            \State $SOM\_tmp\_list \gets child\_process\_info$
            \If {$SOM\_tmp\_list == empty$}
                \State \textbf{break} \Comment{Training is finished}
            \EndIf
        \EndWhile
    \end{algorithmic}
\end{algorithm}

The Vertical Growth Function (VGF) is responsible for calculating the total error of a given SOM object. The total error is then used to calculate the growth threshold criteria. Once these calculations are complete, the VGF compares the error of each neuron in the SOM to the growth threshold. If the error of a given neuron is greater than the growth threshold, then the VGF trains a SOM on that neuron’s information. If the error of a given neuron is not greater than the growth threshold, then the VGF labels that neuron, either benign or malicious, and marks the neuron as a leaf in the HSOM. After analyzing a neuron, the VGF stores the resulting information to send back to the parent process.

\begin{algorithm}
    \label{vgf}
    \caption{Vertical Growth Function}
    \begin{algorithmic}[1]
        \State Calculate the error for each $neuron \in parent\_SOM$
        \State Calculate the \textit{growth\_threshold}
        \For {$neuron \in parent\_SOM$}
            \If {$neuron\_error > growth\_threshold$ \textbf{and} $num\_neuron\_data\_samples > SOM\_GRID\_SIZE$}
                \State Train a SOM on \textit{neuron\_data}
                \State $nodes \gets SOM\_info$
            \Else
                \State Label \textit{neuron} benign or malicious
                \State $nodes \gets neuron$
            \EndIf
        \EndFor
        \State Return \textit{nodes}
    \end{algorithmic}
\end{algorithm}

By detailing how the parHSOM algorithm functions, this work has shown that parHSOM provides a solution to RQ1.

\subsection{Implementation}
This work tested parHSOM on two different testing environments. Testbed 1 was an Alienware Aurora R16 computer, which has 1 socket, 24 CPU cores, and 32 logical processors, and Testbed 2 was a H100 server, which has 2 sockets, 16 CPU cores per socket, and 64 logical processors. Since different output grid sizes can vary the results of a SOM, this research tested each model on four different grid sizes - 2x2, 3x3, 4x4, and 5x5. Previous literature has used accuracy, precision, recall, F1 score, False Positive Rate (FPR), False Negative Rate (FNR), Training Time (TT), and Prediction Time (PT) to evaluate the performance of the HSOM algorithm \cite{ables2023explainable}. Thus, this work used the same metrics to evaluate the success of parHSOM. Each test was conducted ten times and the average result of those tests was collected for each evaluation metric.

Five cybersecurity datasets were used to assess the efficacy of parHSOM in a variety of scenarios: CIC-IDS-2017, CIC-IDS-2018, NSL-KDD, TON\_IoT, and UNSW-NB15. These datasets were selected because they have been used for intrusion detection research in the literature for a variety of research fields, such as Construction \& Improvement, Internet of Things (IoT), Medical, Network, and Vehicular. The Construction \& Improvement field of research encompasses research that focuses on techniques for improving or building IDSs without seeking to improve a specific area of IDS. The IoT, Medical, Network, and Vehicle research fields include methods that specifically focus improving their respective areas. The metatdata for each dataset was compiled into Table \ref{tab_dataset_breakdown}. The following subsections cover the details of each dataset in more depth.

\input{tab_dataset_breakdown}

\subsubsection{CIC-IDS-2017}
The authors of \cite{sharafaldin2018toward} published CIC-IDS-2017 in 2017. The data was collected by the Canadian Institute for Cybersecurity (CIC) at UNB using two completely separated networks - a Victim-Network and an Attack-Network. A benign-profile (B-profile) was created to imitate human interactions on a network, and six attack profiles were created based on an updated list of common attack families.

\subsubsection{CIC-IDS-2018}
Released by CIC in 2018, CIC-IDS-2018 \cite{sharafaldin2018toward} was designed to be an upgraded version of CIC-IDS-2017. The data was collected using two completely separated networks - a Victim-Network and an Attack-Network. A B-profile was created to imitate human interactions on a network, and six attack profiles were used to create malicious traffic.

\subsubsection{NSL-KDD}
The University of New Brunswick (UNB) \cite{tavallaee2009detailed} released NSL-KDD in 2009. It was designed as an updated version of KDDCup99 dataset and has four main types of attacks - DoS, U2R, R2L, and Probing.

\subsubsection{TON\_IoT}
The authors of \cite{moustafa2021new} released TON\_IoT in 2019 by UNSW. This dataset was created by UNSW to address the lack of authoritative and representative heterogeneous information for IoT networks. The data was collected from a three layer architecture, which simulated a realistic implementation of recent real-world IoT/Industrial IoT (IIoT) networks - the edge layer, which contained the physical devices and their operating systems; the fog layer, which contained the virtualization technology that programs and controls the VMs; and the cloud layer, which contains the cloud services configured online in the testbed. TON\_IoT contains nine types of attacks - Scanning, DoS, DDoS, Ransomware, Backdoor, Injection, XXS, Password Cracking, and Man-In-The-Middle (MITM).

\subsubsection{UNSW-NB15}
Created by UNSW and released in 2015 \cite{moustafa2015unsw}, UNSW-NB15 was designed to provide a comprehensive network based dataset. The data was collected from a virtual architecture which consisted of three virtual servers - two normal and one abnormal. This dataset has 197 features and 9 types of attacks - Fuzzers, Analysis, Backdoor, DoS, Exploits, Generic, Reconnaissance, Shellcode, and Worms.

\section{Results} \label{results}
In this section, the results for each set of dataset experiments are provided. As mentioned in Section \ref{design}, the metrics used to gauge the performance of parHSOM and the Sequential HSOM are precision, recall, F1 score, accuracy, False Positive Rate (FPR), False Negative Rate (FNR), Training Time (TT), and Prediction Time (PT). The result averages for the CIC-IDS-2017 dataset experiments on Testbed 1 and Testbed 2 are displayed in Table \ref{tab_cicids2017_computer} and Table \ref{tab_cicids2017_server}. CIC-IDS-2017 was the second largest dataset that was used to test parHSOM. Additionally, it has the lowest contamination rate. The result averages for the CIC-IDS-2018 dataset experiments on Testbed 1 and Testbed 2 are shown in Table \ref{tab_cicids2018_computer} and Table \ref{tab_cicids2018_server}. CIC-IDS-2018 was the largest dataset that parHSOM was tested on, and it had the second smallest contamination rate. Table \ref{tab_nslkdd_computer} and Table \ref{tab_nslkdd_server} provide the average results for the NSL-KDD experiments on Testbed 1 and Testbed 2. NSL-KDD was the smallest dataset used to test parHSOM. It has the third smallest contamination rate. The result averages for the TON\_IoT dataset experiments on Testbed 1 and Testbed 2 are shown in Table \ref{tab_toniot_computer} and Table \ref{tab_toniot_server}. TON\_IoT is the second smallest dataset used in this research, and it has the highest contamination rate. Finally, the result averages for the UNSW-NB15 dataset experiments on Testbed 1 and Testbed 2 are displayed in Table \ref{tab_unswnb15_computer} and Table \ref{tab_unswnb15_server}. UNSW-NB15 is the third smallest dataset used to test parHSOM, and it has the second highest contamination rate.

\input{results/tab_CIC-IDS-2017_computer}
\input{results/tab_CIC-IDS-2017_server}

\input{results/tab_CIC-IDS-2018_computer}
\input{results/tab_CIC-IDS-2018_server}

\input{results/tab_NSL-KDD_computer}
\input{results/tab_NSL-KDD_server}

\input{results/tab_TON_IoT_computer}
\input{results/tab_TON_IoT_server}

\input{results/tab_UNSW-NB15_computer}
\input{results/tab_UNSW-NB15_server}

\subsection{Speedup Calculations}
\begin{figure}[!ht]
    \centering
    \includegraphics[width=1\linewidth]{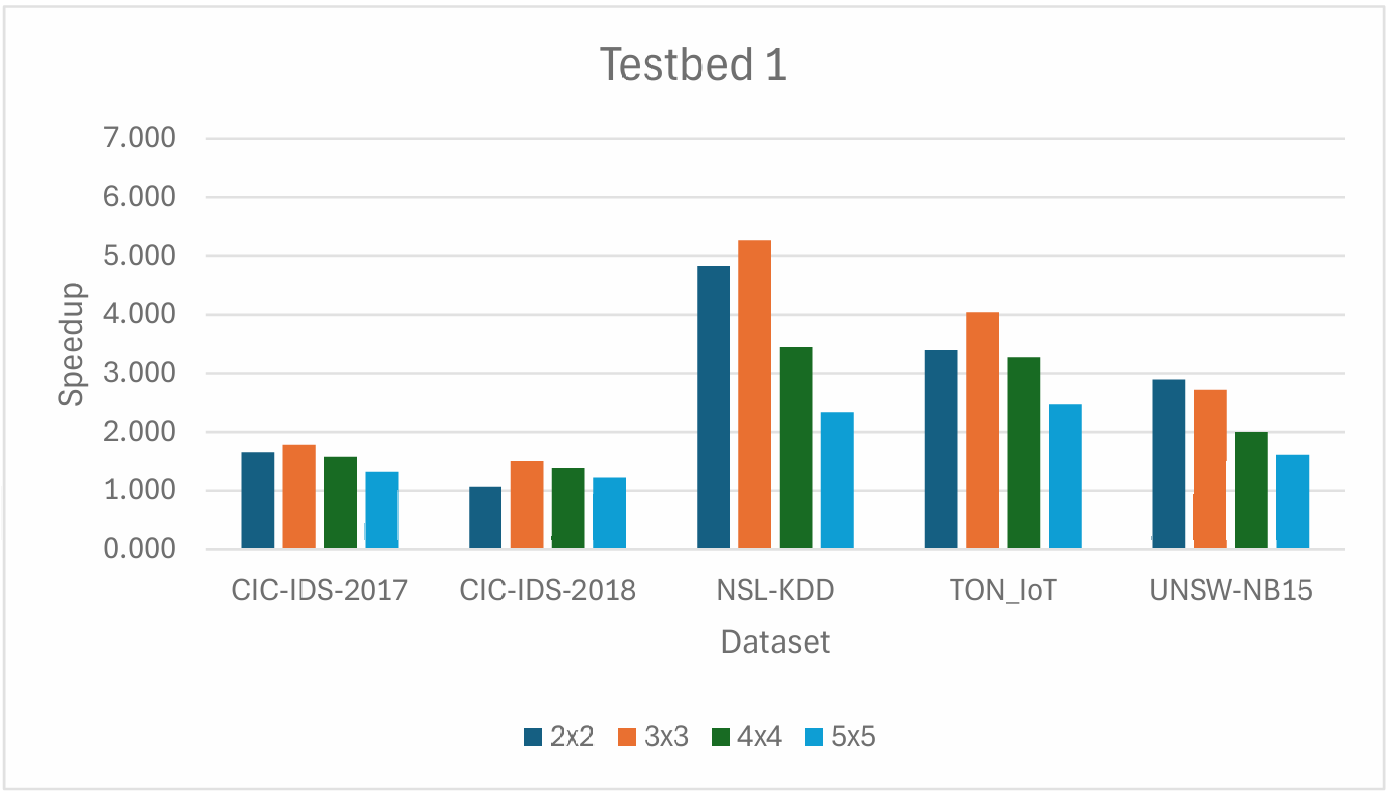}
    \caption{A figure that compares the speed increase of parHSOM across all of the Testbed 1 experiments.}
    \label{fig_speedup_computer}
\end{figure}

\begin{figure}[!ht]
    \centering
    \includegraphics[width=1\linewidth]{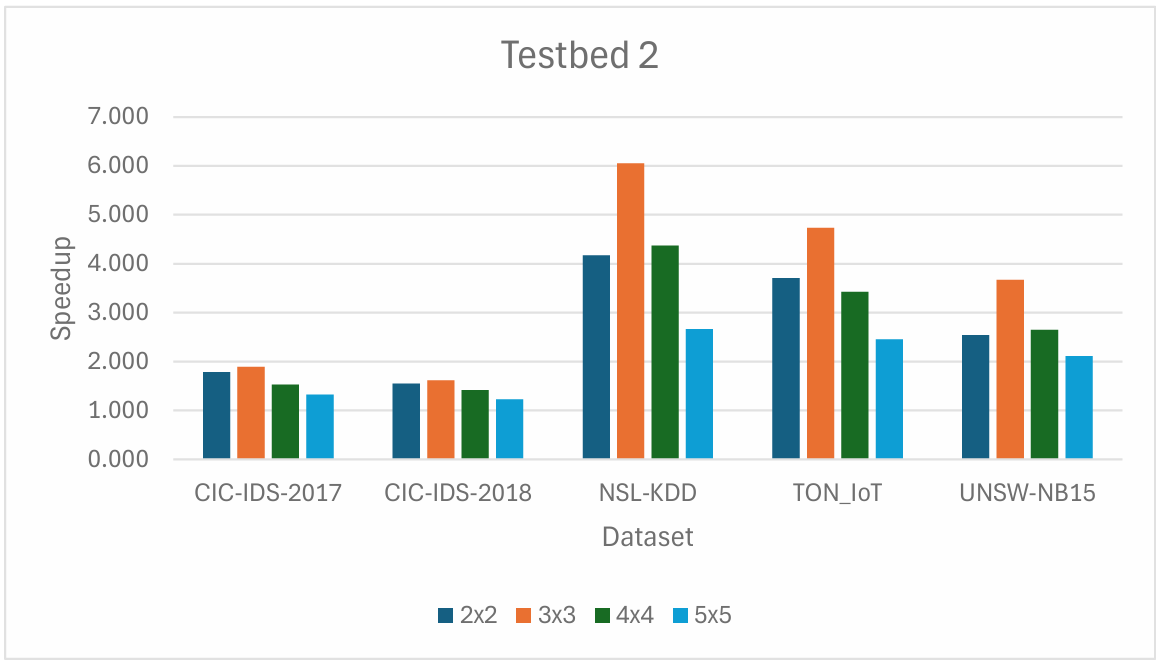}
    \caption{A figure that compares the speed increase of parHSOM across all of the Testbed 2 experiments.}
    \label{fig_speedup_server}
\end{figure}

The speedup for parHSOM was calculated based on the Testbed 1 and Testbed 2 from each dataset's experiment data. The results from each set of experiments were plotted on a bar graph and compared to each other. These comparisons can be found in Figure \ref{fig_speedup_computer} and Figure \ref{fig_speedup_server}. 

\section{Discussion} \label{discussion}
As shown in Section \ref{design}, provides an answer to RQ1, and thus, the focus of the discussion section will be to analyze how the experiment results relate to RQ2.

\subsection{Research Question 2}
To understand how parHSOM algorithm compares to the Sequential HSOM algorithm, this research split RQ2 into two subproblems:
\begin{enumerate}
    \item How does the parHSOM Training Time compare to the Sequential HSOM Training Time?
    \item How do the other parHSOM performance metrics compare to the other Sequential HSOM performance metrics?
\end{enumerate}
RQ2.1 focuses on a central component of this research, which is whether or not parHSOM will train faster than the Sequential HSOM. However, a faster training time is not be useful if the trained parHSOM functions significantly worse than the Sequential HSOM in all of the other performance metrics. Thus, RQ2.2 will focus on how the performance metrics of parHSOM compare to the performance metrics of the Sequential HSOM.

\subsubsection{Research Question 2 - Subproblem 1}
Analyzing the training time results from the previous sections shows that parHSOM consistently outperformed the Sequential HSOM. However, the amount of speedup achieved by parHSOM changed depending on the experiment variables. The best training time improvement for parHSOM happened when it was trained on the NSL-KDD dataset using the 3x3 grid. With this setup, parHSOM achieved a 5.265 speed increase on Testbed 1 and a 6.056 speed increase on Testbed 2. Further analysis of the other dataset experiments showed similar results. As Table \ref{tab_speedup_discussion} shows, parHSOM performed the best using the 3x3 output grid for every set of experiments except for one. The single outlier is when parHSOM was trained on Testbed 1 with the UNSW-NB15 dataset. In that instance, the 2x2 grid size performed the best with a 2.897 speed increase; however, even in that case, the 2x2 output grid only narrowly beat the 3x3 grid, which had a 2.719 speed increase. These observations imply that parHSOM performs best using a 3x3 grid.

\input{results/tab_speedup_discussion}

Another trend this work observed in the training time results was that as the size of the dataset increased, the amount of parHSOM speedup decreased. Since the number of spawned child processes also increases with the size of the dataset, this would imply that as the size of the dataset increases the CPU becomes overloaded with children processes.

While the experiments on Testbed 1 consistently finished training faster than the Testbed 2 experiments, the Testbed 2 experiments showed higher speed increase between parHSOM and the Sequential HSOM. The slower train times on Testbed 2 would imply overhead communication costs between the two sockets. However, the higher speedup implies room for improvement if the socket communication can be optimized.

\subsubsection{Research Question 2 - Subproblem 2}
To assess how parHSOM performed in comparison to the Sequential HSOM, this work calculated the difference between the performance of parHSOM and the Sequential HSOM for all of our experiments. For the CIC-IDS-2017, CIC-IDS-2018, and NSL-KDD experiments, all of the respective performance values were within 0.01 of each other, and for the TON\_IoT and UNSW-NB15 experiments, most of the respective performance values were within 0.01 of each other with a couple of the values being within 0.03 of each. Thus, it can be concluded that no significant change occurred in the precision, recall, F1score, accuracy, FPR, FNR, or PT between the Sequential HSOM and parHSOM. This point is further emphasized when the accuracy, False Positive Rate (FPR), and False Negative Rate (FNR) performance metrics are examined in greater detail. To analyze them further, each group of metrics was converted to percentages and graphed. \newline

\noindent \textbf{Accuracy Metrics:}
First, the accuracy metrics were converted to percentages, and then, the percentages were graphed to compare the outcomes from each of the dataset experiments.
In the CIC-IDS-2017 experiments, all of the accuracy metrics were within 0.2\% of each other, and parHSOM performed the best on Testbed 1 with a 2x2 grid and on Testbed 2 with a 3x3 grid.
For CIC-IDS-2018, the accuracy metrics were within 0.9\% of each other, and parHSOM performed the best on both Testbed 1 and Testbed 2 with a 3x3 grid.
In the NSL-KDD experiments, the accuracy range from 97.00\% to 97.50\%, which results in a difference of 0.5\%. parHSOM had the same level of accuracy for each of the grid sizes on Testbed 1, and it performed the best on Testbed 2 with both the 2x2 and 3x3 grid.
The TON\_IoT experiments showed the largest difference between the accuracy metrics for this research with a range of 1.6\%; however, for each of the different grid sizes, parHSOM performed better than or very similar to the Sequential HSOM. parHSOM performed the best on both Testbed 1 and Testbed 2 with the 4x4 grid. However, it is interesting to note that the largest difference between the Sequential HSOM and parHSOM was for the 2x2 grid experiments where parHSOM performed better than the Sequential HSOM by 1\%.
Finally for UNSW-NB15, the accuracy metrics were within 1.0\% of each other, and parHSOM performed the best on both Testbed 1 and Testbed 2 with the 2x2 grid.

\noindent \textbf{False Positive Rate (FPR) Metrics:}
Similar to the accuracy metrics, the FPR metrics were converted into percentages and graphed in order to analyze the difference between parHSOM and the Sequential HSOM.
For CIC-IDS-2017, the FPR metrics ranged from 2.16\% to 2.68\%, which results in a difference of 0.52\%. parHSOM performed the best on Testbed 1 with the 2x2 and 3x3 grid and on Testbed 2 with the 3x3 grid.
In the CIC-IDS-2018 experiments, all of the FPR metrics were within 0.36\% of each other, and parHSOM performed the best on both Testbed 1 and Testbed 2 with the 3x3 grid.
The NSL-KDD experiments presented FPR metrics that were within 0.31\% of each other, and parHSOM performed the best on both Testbed 1 and Testbed 2 with the 2x2 grid.
The TON\_IoT experiments showed a larger split with the percentages ranging from 4.46\% to 8.57\%, for a difference of 4.11\%. However, parHSOM performed better than the Sequential HSOM for all grid sizes except the 5x5 grid on Testbed 2, and in that case, parHSOM only differed from the Sequential HSOM by 0.70\%. parHSOM performed the best on both Testbed 1 and Testbed 2 with the 4x4 grid size.
Finally, in the UNSW-NB15 experiments, this research recorded the largest gap of percentages, with a range of 7.07\%. However, the larger percentage range is due to the different grid sizes producing different results. Within the same grid size, parHSOM performed similar to, if not better than, the Sequential HSOM. parHSOM performed the best on both Testbed 1 and Testbed 2 with the 2x2 grid.

\noindent \textbf{False Negative Rate (FNR) Metrics:}
Like the previous section, the False Negative Metrics were converted into percentages and then graphed in order to analyze how parHSOM compared to the Sequential HSOM.
In the CIC-IDS-2017 experiments, all of the FNR metrics were within 1.55\% of each other. Analyzing the metrics showed that the largest gap between the Sequential HSOM and parHSOM was in the experiments using the 3x3 grid size on Testbed 1. However, even then, the models only differed by 1.08\%. In all other cases, the difference between the Sequential HSOM and parHSOM was less than 0.5\%, and for the 2x2 grid experiments on Testbed 1, parHSOM outperformed the Sequential HSOM. parHSOM performed the best on both Testbed 1 and Testbed 2 with the 5x5 grid.
The CIC-IDS-2018 experiments showed FNR metrics that were within 1.33\% of each other. This range comes from the various grid sizes performing differently. In all cases, the difference between parHSOM and the Sequential HSOM was less than 0.5\%, in most of the cases, parHSOM differed from the Sequential HSOM by less than 0.25\%, and in four of the cases, parHSOM performed better than the Sequential HSOM. parHSOM performed best on both Testbed 1 and Testbed 2 with the 3x3 grid.
In the NSL-KDD experiments, the FNR metrics ranged from 3.08\% to 4.11\%. In all cases, parHSOM differed from the Sequential HSOM by less than 0.5\%, and in five out of the eight experiment setups, parHSOM performed better than the Sequential HSOM. parHSOM performed the best on both Testbed 1 and Testbed 2 with the 3x3 grid.
In the TON\_IoT experiments, the FNR metrics ranged from 1.26\% to 8.25\%. The experiments on Testbed 1 resulted in consistently higher FNRs. However, in each case, parHSOM performed better than the Sequential HSOM. The experiments run on Testbed 2 consistently returned lower FNRs. Within those results, parHSOM performed better than the Sequential HSOM on the 3x3 and 4x4 grid sizes, very similar on the 5x5 grid, and less than 1\% different on the 2x2 grid. parHSOM performed best on Testbed 1 with the 4x4 grid and on Testbed 2 with the 3x3 grid.
Finally, for UNSW-NB15, the FNR metrics are all within 3.07\% of each other. The experiments run using the 5x5 grid size performed better than the other grid sizes, and thus, increased the range of FNR metric percentages. parHSOM performed the best on both Testbed 1 and Testbed 2 with the 5x5 grid.


\section{Conclusion} \label{conclusion}
This research set out to design a parallel HSOM implementation. After researching various parallelization methods for the SOM, this work proposed the parHSOM algorithm. parHSOM is based on the data-partitioned parallelization method, and it has been tested on two different environments, four different grid sizes, and five different cybersecurity datasets. Using the experiment data, this research answered two research questions. RQ1 focused solely on whether or not parallelizing the HSOM was possible and was answered in Section \ref{design}. RQ2 focused on how parHSOM performed compared to the Sequential HSOM. In Section \ref{results}, this work divided RQ2 into two subproblems. The first subproblem focused on the Training Time of parHSOM, while the second subproblem focused on the other performance metrics of parHSOM. In subproblem 1, this work showed that parHSOM consistently performed better than the Sequential HSOM, with a maximum speed increase of 6.056. Analyzing the other performance metrics provided an answer to subproblem 2, where this work showed that parHSOM performed similarly, if not better than, the Sequential HSOM. Overall, these experiments have shown that parHSOM contains potential as a parallel HSOM implementation. In the next section, this work covers some of the limitations of this research and how those limitations link to areas of future research.

\subsection{Limitations}
Due to the lack of literature on parallel HSOM implementations, this work started with the base level components of the HSOM algorithm and analyzed what happened when those base components were parallelized using the data-partitioned method. However, this research was not without its limitations. First, while the HSOM algorithm was designed to mitigate the SOM algorithm's sensitivity to initial conditions, the HSOM is not immune to this weakness. This work standardized all of the initial conditions, such as the learning rate and neighborhood function, except for the output grid size which was limited to four specific sizes. This decision was made so that the effect of parallelizing the HSOM could be better observed due to the reduced amount of changing factors. However, as the size of the weight updates decreased, the model sent Runtime error messages indicating that the initialization factors could be better optimized. Another limitation of this work is the fact that parHSOM was designed using Python. Python was chosen to help provide a starting point for parallel HSOM research, but it is not the optimal programming language for reducing communication overhead. Instead of being drawbacks, these limitation provide insight into future areas of research, which will be discussed in the next section.

\subsection{Future Work}
parHSOM shows that there is potential in parallelizing the HSOM algorithm. One area of future research would be to analyze what happens if parHSOM is implemented in a more efficient programming language, such as MPI. With this language, researchers could investigate whether or not the communication overhead of parHSOM can be decreased. Another area of future research would be to investigate how initialization parameters affect parHSOM's performance. parHSOM could be expanded into a Growing Hierarchical Self-Organizing Map (GHSOM), which has been shown in the literature to help further reduce the SOM algorithm's sensitivity to initial conditions. Finally, a third area of future research would be to investigate how parHSOM performs if it was modified to utilize GPU resources.

In conclusion, this research provides a starting point for parallel implementations of the HSOM algorithm Furthermore, it highlights areas of future research. Finally, it demonstrates that parallelizing the HSOM algorithm is a promising field of research for improving explainable, trustworthy, and AI-based IDSs.

\section{Acknowledgments}
The authors would like to acknowledge the support of the Predictive Analytics and Technology Integration (PATENT) Laboratory for their support regarding finances, technology, and leadership in this work.

\bibliographystyle{IEEEtran}
\bibliography{references}
\end{document}

%% file: tab_dataset_breakdown.tex
\begin{table*}[!ht]
    \resizebox{\textwidth}{!}{
    {\rowcolors{3}{gray!20}{white}
    \begin{tabular}{lc |*{3}{c} |*{8}{c}| *{1}{c}}
    \hline\hline
    \multicolumn{2}{c}{\textbf{}} &
        \multicolumn{3}{c}{\textbf{Metadata}} &
        \multicolumn{8}{c}{\textbf{Attack Categories}} \\ \hline
    \rule{0pt}{20mm}    
    \textbf{Datasets} &
        \textbf{Year} &
        \rotatebox[origin=c]{90}{\parbox{35mm}{\centering \textbf{Number of Samples}}} &
        \rotatebox[origin=c]{90}{\parbox{35mm}{\centering \textbf{Contamination Rate}}} &
        \rotatebox[origin=c]{90}{\parbox{35mm}{\centering \textbf{Number of Features}}} &
        \rotatebox[origin=c]{90}{\parbox{35mm}{\centering \textbf{Botnet}}} &
        \rotatebox[origin=c]{90}{\parbox{35mm}{\centering \textbf{Brute Force}}} &
        \rotatebox[origin=c]{90}{\parbox{35mm}{\centering \textbf{DoS / DDoS}}} &
        \rotatebox[origin=c]{90}{\parbox{35mm}{\centering \textbf{Escalation}}} &
        \rotatebox[origin=c]{90}{\parbox{35mm}{\centering \textbf{Infiltration}}} &
        \rotatebox[origin=c]{90}{\parbox{35mm}{\centering \textbf{Injection}}} &
        \rotatebox[origin=c]{90}{\parbox{35mm}{\centering \textbf{Malware}}} &
        \rotatebox[origin=c]{90}{\centering\parbox{35mm}{\centering \textbf{Reconnaissance}}} &
        \textbf{\parbox{25mm}{\centering IDS Research Categories}} \\ \hline
    \parbox{30mm}{\raggedright \textbf{NSL-KDD} \cite{tavallaee2009detailed}}        
        & 2009 & 148,517 & 48.12\% & 122 & - & - & \checkmark & \checkmark & \checkmark & - & - & \checkmark & \parbox{70mm}{\raggedright Construction \& Improvement, IoT, Medical, Network, Vehicular} \\ 
    \parbox{30mm}{\raggedright \textbf{UNSW-NB15} \cite{moustafa2015unsw}}      
        & 2015 & 257,673   & 63.91\% & 197 & - & - & \checkmark & \checkmark & \checkmark & - & \checkmark & \checkmark & \parbox{70mm}{\raggedright Construction \& Improvement, IoT, Medical, Network, Vehicular} \\
    \parbox{30mm}{\raggedright \textbf{CIC-IDS-2017} \cite{sharafaldin2018toward}}    
        & 2017 & 2,827,876         & 19.68\% & 78 & \checkmark & \checkmark & \checkmark & - & \checkmark & \checkmark & - & \checkmark & \parbox{70mm}{\raggedright Construction \& Improvement, Network, Vehicular} \\
    \parbox{30mm}{\raggedright \textbf{CIC-IDS-2018} \cite{sharafaldin2018toward}}   
        & 2018 & 7,199,312*         & 20.60\% & 81 & \checkmark & \checkmark & \checkmark & - & \checkmark & - & - & - & \parbox{70mm}{\raggedright Construction \& Improvement} \\
    \parbox{30mm}{\raggedright \textbf{TON-IoT} \cite{moustafa2021new}}         
        & 2019 & 211,042         & 76.31\% & 82 & - & \checkmark & \checkmark & - & \checkmark & \checkmark & \checkmark & - & \parbox{70mm}{\raggedright Construction \& Improvement, IoT} \\
    \hline\hline
    \end{tabular}}
    }
    \caption{A table that lists the details of the cybersecurity datasets used during the experiments for this research.}
    \label{tab_dataset_breakdown}
\end{table*}

%% file: results/tab_CIC-IDS-2017_computer.tex
\begin{table}
    \resizebox{0.5\textwidth}{!}{
\begin{tabular}{clcccccc}
    \hline\hline
    \multicolumn{8}{c}{} \\
    \multicolumn{8}{c}{\multirow{-2}{*}{\textbf{CIC-IDS-2017 Dataset: HSOM Training Results}}} \\
    \hline\hline
    \multicolumn{8}{c}{\textbf{Testbed 1}} \\
    \bottomrule
    \multicolumn{2}{c}{}                                      & \multicolumn{2}{c}{Precision}                                   & \multicolumn{2}{c}{Recall}                                        & \multicolumn{2}{c}{F1 Score}                                  \\
    \multicolumn{2}{c}{\multirow{-2}{*}{\textbf{Sequential}}} & 0                             & 1                               & 0                               & 1                               & 0                             & 1                             \\ \midrule
                                & \grey 2x2     & \grey 0.987   & \grey 0.907   & \grey 0.976   & \grey 0.941   & \grey 0.980   & \grey 0.925   \\
                                & 3x3           & 0.990         & 0.912         & 0.978         & 0.957         & 0.980         & 0.932         \\
                                & \grey 4x4     & \grey 0.989   & \grey 0.907   & \grey 0.977   & \grey 0.948   & \grey 0.980   & \grey 0.927   \\
\bottomrule
    \end{tabular}
    }
    \caption{A table that contains the result averages for the experiments conducted using the CIC-IDS-2017 dataset on Testbed 1.}
    \label{tab_cicids2017_computer}
\end{table}

%% file: results/tab_CIC-IDS-2017_server.tex
\begin{table}
    \resizebox{0.5\textwidth}{!}{
    \begin{tabular}{clcccccc}
    \hline\hline
    \multicolumn{8}{c}{} \\
    \multicolumn{8}{c}{\multirow{-2}{*}{\textbf{CIC-IDS-2017 Dataset: HSOM Training Results (continued)}}} \\
    \hline\hline
    \multicolumn{8}{c}{\textbf{Testbed 2}} \\
    \bottomrule
    \multicolumn{2}{c}{}                                      & \multicolumn{2}{c}{Precision}                                   & \multicolumn{2}{c}{Recall}                                        & \multicolumn{2}{c}{F1 Score}                                  \\
    \multicolumn{2}{c}{\multirow{-2}{*}{\textbf{Sequential}}} & 0                             & 1                               & 0                               & 1                               & 0                             & 1                             \\ \midrule
                                & \grey 2x2     & \grey 0.987   & \grey 0.902   & \grey 0.976   & \grey 0.942   & \grey 0.979   & \grey 0.922   \\
                                & 3x3           & 0.990         & 0.905         & 0.977         & 0.950         & 0.980         & 0.927         \\
                                & \grey 4x4     & \grey 0.989   & \grey 0.909   & \grey 0.979   & \grey 0.944   & \grey 0.980   & \grey 0.928   \\
\bottomrule
    \end{tabular}
    }
    \caption{A table that contains the result averages for the experiments conducted using the CIC-IDS-2017 dataset on Testbed 2.}
    \label{tab_cicids2017_server}
\end{table}

%% file: results/tab_CIC-IDS-2018_computer.tex
\begin{table}
    \resizebox{0.5\textwidth}{!}{
    \begin{tabular}{clcccccc}
    \hline\hline
    \multicolumn{8}{c}{} \\
    \multicolumn{8}{c}{\multirow{-2}{*}{\textbf{CIC-IDS-2018 Dataset: HSOM Training Results}}} \\
    \hline\hline
    \multicolumn{8}{c}{\textbf{Testbed 1}} \\
    \bottomrule
    \multicolumn{2}{c}{}                                      & \multicolumn{2}{c}{Precision}                                   & \multicolumn{2}{c}{Recall}                                        & \multicolumn{2}{c}{F1 Score}                                  \\
    \multicolumn{2}{c}{\multirow{-2}{*}{\textbf{Sequential}}} & 0                             & 1                               & 0                               & 1                               & 0                             & 1                             \\ \midrule
                                & \grey 2x2     & \grey 0.965   & \grey 0.962   & \grey 0.990   & \grey 0.859   & \grey 0.980   & \grey 0.907   \\
                                & 3x3           & 0.969         & 0.963         & 0.990         & 0.867         & 0.980         & 0.911         \\
                                & \grey 4x4     & \grey 0.967   & \grey 0.967   & \grey 0.990   & \grey 0.863   & \grey 0.980   & \grey 0.911   \\
\bottomrule
    \end{tabular}
    }
    \caption{A table that contains the result averages for the experiments conducted using the CIC-IDS-2018 dataset on Testbed 1.}
    \label{tab_cicids2018_computer}
\end{table}

%% file: results/tab_CIC-IDS-2018_server.tex
\begin{table}
    \resizebox{0.5\textwidth}{!}{
    \begin{tabular}{clcccccc}
    \hline\hline
    \multicolumn{8}{c}{} \\
    \multicolumn{8}{c}{\multirow{-2}{*}{\textbf{CIC-IDS-2018 Dataset: HSOM Training Results (continued)}}} \\
    \hline\hline
    \multicolumn{8}{c}{\textbf{Testbed 2}} \\
    \bottomrule
    \multicolumn{2}{c}{}                                      & \multicolumn{2}{c}{Precision}                                   & \multicolumn{2}{c}{Recall}                                        & \multicolumn{2}{c}{F1 Score}                                  \\
    \multicolumn{2}{c}{\multirow{-2}{*}{\textbf{Sequential}}} & 0                             & 1                               & 0                               & 1                               & 0                             & 1                             \\ \midrule
                                & \grey 2x2     & \grey 0.966   & \grey 0.961   & \grey 0.990   & \grey 0.859   & \grey 0.979   & \grey 0.906   \\
                                & 3x3           & 0.968         & 0.965         & 0.990         & 0.864         & 0.980         & 0.911         \\
                                & \grey 4x4     & \grey 0.968   & \grey 0.964   & \grey 0.990   & \grey 0.863   & \grey 0.980   & \grey 0.909   \\
\bottomrule
    \end{tabular}
    }
    \caption{A table that contains the result averages for the experiments conducted using the CIC-IDS-2018 dataset on Testbed 2.}
    \label{tab_cicids2018_server}
\end{table}

%% file: results/tab_NSL-KDD_computer.tex
\begin{table}
    \resizebox{0.5\textwidth}{!}{
    \begin{tabular}{clcccccc}
    \hline\hline
    \multicolumn{8}{c}{} \\
    \multicolumn{8}{c}{\multirow{-2}{*}{\textbf{NSL-KDD Dataset: Training HSOM Results}}} \\
    \hline\hline
    \multicolumn{8}{c}{\textbf{Testbed 1}} \\
    \bottomrule
    \multicolumn{2}{c}{}                                      & \multicolumn{2}{c}{Precision}                                   & \multicolumn{2}{c}{Recall}                                        & \multicolumn{2}{c}{F1 Score}                                  \\
    \multicolumn{2}{c}{\multirow{-2}{*}{\textbf{Sequential}}} & 0                             & 1                               & 0                               & 1                               & 0                             & 1                             \\ \midrule
                                & \grey2x2 & \grey0.968 & \grey0.977   & \grey0.980   & \grey0.962   & \grey0.975 & \grey0.971 \\
                                & 3x3                         & 0.971                         & 0.976                           & 0.979                           & 0.968                           & 0.976                         & 0.971                         \\
                                & \grey4x4 & \grey0.969 & \grey0.976   & \grey0.979   & \grey0.966   & \grey0.973 & \grey0.971 \\
\bottomrule
    \end{tabular}
    }
    \caption{A table that contains the result averages for the experiments conducted using the NSL-KDD dataset on Testbed 1.}
    \label{tab_nslkdd_computer}
\end{table}

%% file: results/tab_NSL-KDD_server.tex
\begin{table}
    \resizebox{0.5\textwidth}{!}{
    \begin{tabular}{clcccccc}
    \hline\hline
    \multicolumn{8}{c}{} \\
    \multicolumn{8}{c}{\multirow{-2}{*}{\textbf{NSL-KDD Dataset: HSOM Training Results (continued)}}} \\
    \hline\hline
    \multicolumn{8}{c}{\textbf{Testbed 2}} \\
    \bottomrule
    \multicolumn{2}{c}{}                                      & \multicolumn{2}{c}{Precision}                                   & \multicolumn{2}{c}{Recall}                                        & \multicolumn{2}{c}{F1 Score}                                  \\
    \multicolumn{2}{c}{\multirow{-2}{*}{\textbf{Sequential}}} & 0                             & 1                               & 0                               & 1                               & 0                             & 1                             \\ \midrule
                                & \grey 2x2     & \grey 0.969   & \grey 0.977   & \grey 0.978   & \grey 0.962   & \grey 0.974   & \grey 0.969   \\
                                & 3x3           & 0.971         & 0.977         & 0.979         & 0.969         & 0.973         & 0.971         \\
                                & \grey 4x4     & \grey 0.967   & \grey 0.976   & \grey 0.977   & \grey 0.965   & \grey 0.970   & \grey 0.970   \\
\bottomrule
    \end{tabular}
    }
    \caption{A table that contains the result averages for the experiments conducted using the NSL-KDD dataset on Testbed 2.}
    \label{tab_nslkdd_server}
\end{table}

%% file: results/tab_TON_IoT_computer.tex
\begin{table}
    \resizebox{0.5\textwidth}{!}{
    \begin{tabular}{clcccccc}
    \hline\hline
    \multicolumn{8}{c}{} \\
    \multicolumn{8}{c}{\multirow{-2}{*}{\textbf{TON\_IoT Dataset: HSOM Training Results}}} \\
    \hline\hline
    \multicolumn{8}{c}{\textbf{Testbed 1}} \\
    \bottomrule
    \multicolumn{2}{c}{}                                      & \multicolumn{2}{c}{Precision}                                   & \multicolumn{2}{c}{Recall}                                        & \multicolumn{2}{c}{F1 Score}                                  \\
    \multicolumn{2}{c}{\multirow{-2}{*}{\textbf{Sequential}}} & 0                             & 1                               & 0                               & 1                               & 0                             & 1                             \\ \midrule
                                & \grey 2x2     & \grey 0.938   & \grey 0.974   & \grey 0.918   & \grey 0.981   & \grey 0.927   & \grey 0.976   \\
                                & 3x3           & 0.951         & 0.983         & 0.942         & 0.986         & 0.946         & 0.985         \\
                                & \grey 4x4     & \grey 0.956   & \grey 0.984   & \grey 0.944   & \grey 0.987   & \grey 0.950   & \grey 0.987   \\
\bottomrule
    \end{tabular}
    }
    \caption{A table that contains the result averages for the experiments conducted using the TON\_IoT dataset on Testbed 1.}
    \label{tab_toniot_computer}
\end{table}

%% file: results/tab_TON_IoT_server.tex
\begin{table}
    \resizebox{0.5\textwidth}{!}{
    \begin{tabular}{clcccccc}
    \hline\hline
    \multicolumn{8}{c}{} \\
    \multicolumn{8}{c}{\multirow{-2}{*}{\textbf{TON\_IoT Dataset: HSOM Training Results (continued)}}} \\
    \hline\hline
    \multicolumn{8}{c}{\textbf{Testbed 2}} \\
    \bottomrule
    \multicolumn{2}{c}{}                                      & \multicolumn{2}{c}{Precision}                                   & \multicolumn{2}{c}{Recall}                                        & \multicolumn{2}{c}{F1 Score}                                  \\
    \multicolumn{2}{c}{\multirow{-2}{*}{\textbf{Sequential}}} & 0                             & 1                               & 0                               & 1                               & 0                             & 1                             \\ \midrule
                                & \grey 2x2     & \grey 0.941   & \grey 0.973   & \grey 0.914   & \grey 0.982   & \grey 0.927   & \grey 0.976   \\
                                & 3x3           & 0.956         & 0.986         & 0.946         & 0.988         & 0.950         & 0.985         \\
                                & \grey 4x4     & \grey 0.952   & \grey 0.986   & \grey 0.955   & \grey 0.985   & \grey 0.955   & \grey 0.986   \\
\bottomrule
    \end{tabular}
    }
    \caption{A table that contains the result averages for the experiments conducted using the TON\_IoT dataset on Testbed 2.}
    \label{tab_toniot_server}
\end{table}

%% file: results/tab_UNSW-NB15_computer.tex
\begin{table}
    \resizebox{0.5\textwidth}{!}{
    \begin{tabular}{clcccccc}
    \hline\hline
    \multicolumn{8}{c}{} \\
    \multicolumn{8}{c}{\multirow{-2}{*}{\textbf{UNSW-NB15 Dataset: HSOM Training Results}}} \\
    \hline\hline
    \multicolumn{8}{c}{\textbf{Testbed 1}} \\
    \bottomrule
    \multicolumn{2}{c}{}                                      & \multicolumn{2}{c}{Precision}                                   & \multicolumn{2}{c}{Recall}                                        & \multicolumn{2}{c}{F1 Score}                                  \\
    \multicolumn{2}{c}{\multirow{-2}{*}{\textbf{Sequential}}} & 0                             & 1                               & 0                               & 1                               & 0                             & 1                             \\ \midrule
                                & \grey 2x2     & \grey 0.899   & \grey 0.932   & \grey 0.881   & \grey 0.944   & \grey 0.889   & \grey 0.940   \\
                                & 3x3           & 0.894         & 0.927         & 0.870         & 0.940         & 0.883         & 0.936         \\
                                & \grey 4x4     & \grey 0.886   & \grey 0.928   & \grey 0.871   & \grey 0.936   & \grey 0.879   & \grey 0.930   \\
\bottomrule
    \end{tabular}
    }
    \caption{A table that contains the result averages for the experiments conducted using the UNSW-NB15 dataset on Testbed 1.}
    \label{tab_unswnb15_computer}
\end{table}

%% file: results/tab_UNSW-NB15_server.tex
\begin{table}
    \resizebox{0.5\textwidth}{!}{
    \begin{tabular}{clcccccc}
    \hline\hline
    \multicolumn{8}{c}{} \\
    \multicolumn{8}{c}{\multirow{-2}{*}{\textbf{UNSW-NB15 Dataset: HSOM Training Results (continued)}}} \\
    \hline\hline
    \multicolumn{8}{c}{\textbf{Testbed 2}} \\
    \bottomrule
    \multicolumn{2}{c}{}                                      & \multicolumn{2}{c}{Precision}                                   & \multicolumn{2}{c}{Recall}                                        & \multicolumn{2}{c}{F1 Score}                                  \\
    \multicolumn{2}{c}{\multirow{-2}{*}{\textbf{Sequential}}} & 0                             & 1                               & 0                               & 1                               & 0                             & 1                             \\ \midrule
                                & \grey 2x2     & \grey 0.898   & \grey 0.930   & \grey 0.875   & \grey 0.946   & \grey 0.887   & \grey 0.938   \\
                                & 3x3           & 0.896         & 0.927         & 0.871         & 0.942         & 0.885         & 0.937         \\
                                & \grey 4x4     & \grey 0.884   & \grey 0.929   & \grey 0.873   & \grey 0.935   & \grey 0.880   & \grey 0.930   \\
\bottomrule
    \end{tabular}
    }
    \caption{A table that contains the result averages for the experiments conducted using the UNSW-NB15 dataset on Testbed 2.}
    \label{tab_unswnb15_server}
\end{table}

%% file: results/tab_speedup_discussion.tex
\begin{table}
    \begin{tabular}{ccccc}
    \multicolumn{1}{l}{}                                & \multicolumn{4}{c}{Best parHSOM Training   Time Results}                                                                      \\ \cline{2-5} 
    \multicolumn{1}{l}{}                                & \multicolumn{1}{l}{Speedup}   & \multicolumn{1}{l}{Grid Size} & \multicolumn{1}{l}{Testbed 1} & \multicolumn{1}{l}{Testbed 2} \\ \hline
    \multicolumn{1}{c|}{}                               & 1.778                         & 3x3                           & X                             & -                             \\
    \multicolumn{1}{c|}{\multirow{-2}{*}{CIC-IDS-2017}} & \cellcolor[HTML]{E8E8E8}1.891 & \cellcolor[HTML]{E8E8E8}3x3   & \cellcolor[HTML]{E8E8E8}-     & \cellcolor[HTML]{E8E8E8}X     \\ \cline{1-1}
    \multicolumn{1}{c|}{}                               & 1.501                         & 3x3                           & X                             & -                             \\
    \multicolumn{1}{c|}{\multirow{-2}{*}{CIC-IDS-2018}} & \cellcolor[HTML]{E8E8E8}1.616 & \cellcolor[HTML]{E8E8E8}3x3   & \cellcolor[HTML]{E8E8E8}-     & \cellcolor[HTML]{E8E8E8}X     \\ \cline{1-1}
    \multicolumn{1}{c|}{}                               & 5.265                         & 3x3                           & X                             & -                             \\
    \multicolumn{1}{c|}{\multirow{-2}{*}{NSL-KDD}}      & \cellcolor[HTML]{E8E8E8}6.056 & \cellcolor[HTML]{E8E8E8}3x3   & \cellcolor[HTML]{E8E8E8}-     & \cellcolor[HTML]{E8E8E8}X     \\ \cline{1-1}
    \multicolumn{1}{c|}{}                               & 4.04                          & 3x3                           & X                             & -                             \\
    \multicolumn{1}{c|}{\multirow{-2}{*}{TON\_IoT}}     & \cellcolor[HTML]{E8E8E8}4.734 & \cellcolor[HTML]{E8E8E8}3x3   & \cellcolor[HTML]{E8E8E8}-     & \cellcolor[HTML]{E8E8E8}X     \\ \cline{1-1}
    \multicolumn{1}{c|}{}                               & 2.897                         & 2x2                           & X                             & -                             \\
    \multicolumn{1}{c|}{\multirow{-2}{*}{UNSW-NB15}}    & \cellcolor[HTML]{E8E8E8}3.669 & \cellcolor[HTML]{E8E8E8}3x3   & \cellcolor[HTML]{E8E8E8}-     & \cellcolor[HTML]{E8E8E8}X     \\ \hline
    \end{tabular}
    \caption{A table that shows the conditions for parHSOM's greatest speed increase in each of the dataset experiments.}
    \label{tab_speedup_discussion}
\end{table}